\documentclass{article}

\usepackage{arxiv}

\usepackage[utf8]{inputenc} % allow utf-8 input
\usepackage[T1]{fontenc}    % use 8-bit T1 fonts
\usepackage{hyperref}       % hyperlinks
\usepackage{url}            % simple URL typesetting
\usepackage{booktabs}       % professional-quality tables
\usepackage{amsfonts}       % blackboard math symbols
\usepackage{nicefrac}       % compact symbols for 1/2, etc.
\usepackage{microtype}      % microtypography
\usepackage{lipsum}		% Can be removed after putting your text content
\usepackage{graphicx}
\usepackage{doi}
\usepackage{amsmath}

\title{An Efficient Inference Frame for Single Molecule Localization Microscopy}

\author{ {\includegraphics[scale=0.06]{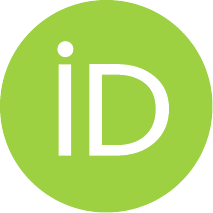}\hspace{1mm}Tingdan Luo}\\
	Department of Biomedical Engineering\\
	Southern University of Science and Technology\\
	\texttt{12132638@mail.sustech.edu.cn} \\
}

\renewcommand{\shorttitle}{An Efficient Inference Frame for Single Molecule Localization Microscopy}

\hypersetup{
pdftitle={An Efficiency Inference Frame for Single Molecule Localization Microscopy},
pdfsubject={q-bio.NC, q-bio.QM},
pdfauthor={TingdanLuo},
pdfkeywords={Parallelization, Light weight, High throughput, SMLM},
}

\begin{document}
\maketitle

\begin{abstract}
Single-molecule localization microscopy (SMLM) surpasses the diffraction limit, achieving subcellular resolution. Traditional SMLM analysis methods often rely on point spread function (PSF) model fitting, limiting the application of complex PSF models. In recent years, deep learning approaches have significantly improved SMLM algorithms, yielding promising results. However, limitations in inference speed and model size have restricted the widespread adoption of deep learning in practical applications. To address these challenges, this paper proposes an efficient model deployment framework and introduces a lightweight neural network, DilatedLoc, aimed at enhancing both image reconstruction quality and inference speed. Compared to leading network models, DilatedLoc reduces network parameters to under 100 MB and achieves a 50\% improvement in inference speed, with superior GPU utilization through a novel deployment architecture compatible with various network models.
\end{abstract}

% keywords can be removed

\section{Introduction}

Deep learning has significantly transformed SMLM data analysis, enabling the extraction of localization information from diffraction-limited images \cite{nehme2018deep, zhang2018analyzing, boyd2018deeploco, zelger2018three, speiser2021deep, nehme2020deepstorm3d}, the reconstruction of high-quality super-resolution images from sparse localization \cite{ouyang2018deep, gaire2020accelerating}, and the optimization of PSF models for specific scenarios \cite{hershko2019multicolor, nehme2021learning}. Despite its broad applications in single-molecule imaging, the high resource demands required for deep learning models have limited their application in biological samples. Therefore, this paper aims to construct a localization network that minimizes parameters while maintaining accuracy, thereby improving inference efficiency.

Based on this, we propose an efficient inference framework for SMLM that significantly improves GPU utilization and inference speed. Current mainstream networks typically suffer from large parameter sizes, high resource consumption, and slow inference times. To address these challenges, we developed a lightweight network with weights under 100 MB, improving inference speed by 50\% compared to FD DeepLoc \cite{fu2023field}.

\section{Efficient Inference Frame}
\subsection{Overview}

Deep learning has been extensively applied in various aspects of super-resolution microscopy. Sparse reconstruction using deep learning minimizes phototoxicity, enabling super-resolution imaging in live cells \cite{ouyang2018deep, gaire2020accelerating}. It also addresses the limitations of traditional methods in high-density localization, improving accuracy \cite{fu2023field, speiser2021deep, nehme2020deepstorm3d}. Deep learning-based methods are also beneficial for extracting hidden molecular information, such as axial position and emission color, enhancing both localization and spectral accuracy. Furthermore, deep learning aids PSF optimization, combining physical models with deep learning for enhanced interpretability.

Despite these advances, most research focuses on improving localization accuracy, overlooking the importance of inference frameworks. The recent rise of large language models (LLMs) has spurred the development of mature inference frameworks such as VLLM\cite{kwon2023efficientmemorymanagementlarge}, Triton, and TGI, which perform exceptionally well in generative AI. However, these frameworks are not suited for SMLM. Beyond network inference speed, the choice of inference framework is also crucial. Therefore, this paper proposes an inference framework specifically designed for SMLM.

\subsection{The Structure of the Frame}
Traditional parallel processing methods, such as PyTorch's data parallelism, enable multi-GPU inference but often fail to fully utilize GPU resources. This limitation arises because single-molecule localization microscopy (SMLM) images are large, while the network inference time is relatively short. Consequently, the speed at which images are loaded into memory is slower than the network’s inference speed, leading to suboptimal GPU utilization. To address this, we have developed an efficient inference framework specifically tailored to the unique demands of SMLM. This framework ensures high GPU utilization even with lightweight network architectures, thereby significantly improving inference speed.

Our proposed inference framework is primarily based on a pipeline design, where multiple processes handle data loading and network inference simultaneously. Although this approach aligns with traditional parallel frameworks, the distinct nature of SMLM data, along with the unique pre- and post-processing requirements for SMLM networks, renders conventional frameworks inadequate in this domain.

\begin{figure}[htbp]
\centering
\includegraphics[width=0.75\linewidth]{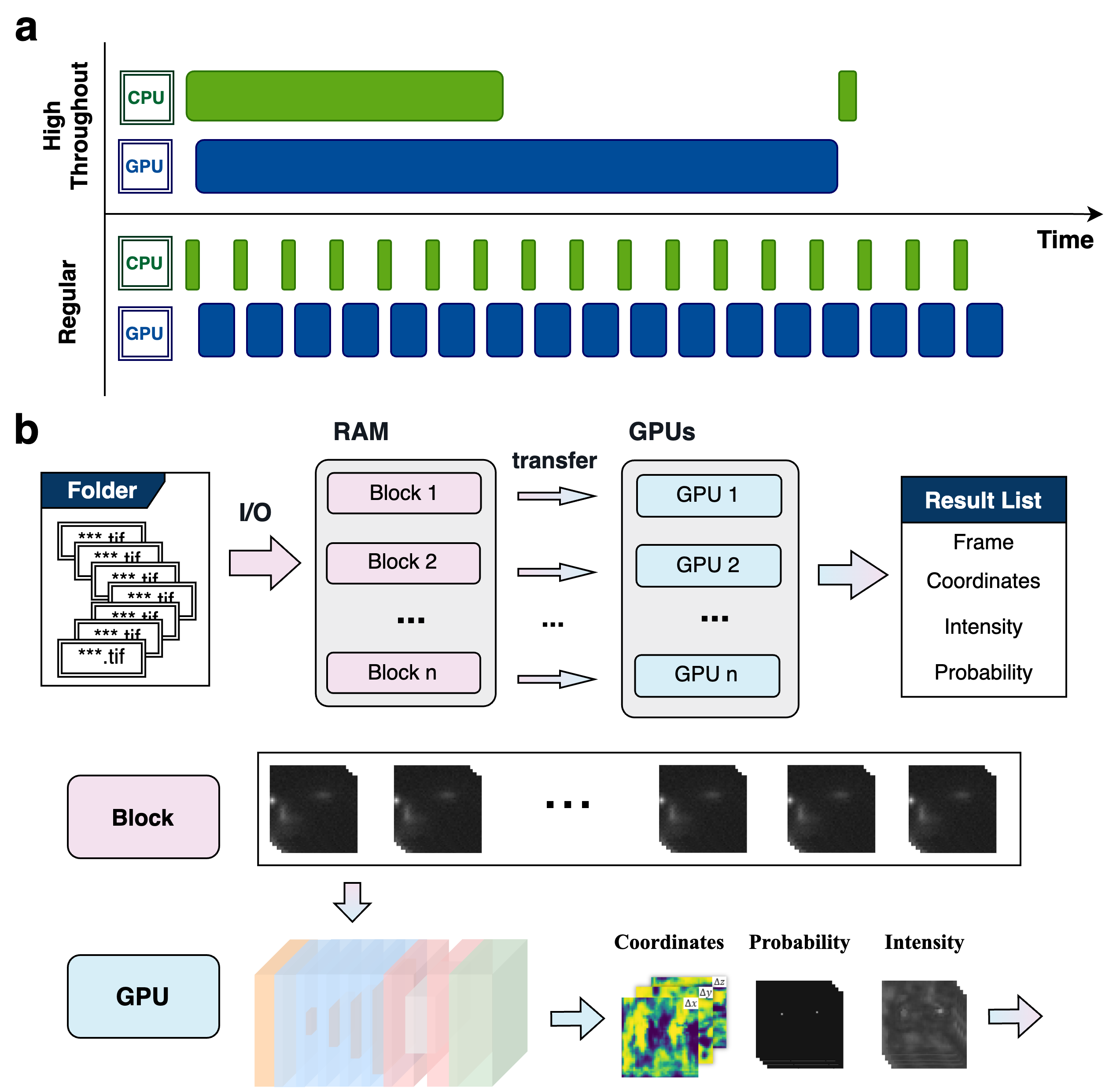}
\caption{Diagram of the reasoning framework. a.CPU, GPU occupancy time for traditional inference vs CPU, GPU occupancy time for high-throughput parallel inference.b }
\label{main}
\end{figure}

The typical processing workflow for a single-molecule image begins by loading the image from disk into CPU memory. From there, the image is copied into GPU memory, where the model performs inference. Once the inference is complete, the results are transferred back to CPU memory for post-processing. Finally, the post-processed coordinates are written to an output file. This entire memory exchange process is illustrated in Figure \ref{main}a.

To optimize this workflow, we made several key improvements. First, post-processing is consolidated within GPU memory to minimize data transfer and reduce copying time. A significant bottleneck in GPU utilization stems from the mismatch between data loading and image inference times. To mitigate this, we implemented a preloading strategy: images are loaded into memory in advance, allowing the network to directly access them from specified memory locations during inference. This optimization ensures smoother data flow and higher overall efficiency. The optimized process flow is depicted in Figure \ref{main}b.

\section{DilatedLoc}
\subsection{Overview}

The challenge of high-density localization is that overlapping PSF shapes at different z-axis positions, under certain background noise, can severely affect the recognition efficiency of fluorescent molecules and significantly impact localization accuracy. High-resolution dense prediction tasks require modeling both local and global patterns in the input domain, as local and global structural features depend on each other. Therefore, simultaneous modeling of local and global structures is crucial for network training. High-density localization is essentially an image feature extraction problem, where convolutional neural networks (CNNs) are most widely used. However, current CNN-based methods perform feature compensation at different resolutions and cannot model both local and global information simultaneously, such as the widely used U-Net \cite{ronneberger2015u}.
Currently, high-density localization networks that perform well typically feature deeper network layers. For example, smNet \cite{zhang2018analyzing}, built on ResNet \cite{he2016deep}, modifies the output channel to extract PSF information. Additionally, upsampling is used to increase image resolution and recover lost details, improving localization accuracy. For instance, DeepStorm3D expands the receptive field with dilated convolution and uses upsampling to enhance localization accuracy.

Due to the characteristics of fluorescent dyes (blinking frequency, blinking duration), the same fluorescent molecule may appear in consecutive frames with different photon intensities. Therefore, compensating for contextual information improves localization accuracy. For example, DECODE \cite{speiser2021deep} improves localization accuracy using contextual three-frame information. However, current networks face challenges such as long training times, large parameter sizes, and difficulty in training, preventing their widespread application. Based on this, this paper proposes DilatedLoc \ref{structure}, a lightweight network suitable for high-density overlapping PSFs with fewer parameters.

\subsection{The Structure of DilatedLoc}

To address the challenges posed by high-density overlap in single-molecule localization, this paper introduces DilatedLoc, a network structure designed to handle overlapping PSFs. DilatedLoc leverages Dense Connection \cite{jegou2017one}, which reduces the difficulty of network training. We propose the Dense Connection Dilated Block (DCD), which combines dense connections with dilated convolutions to enhance feature collection efficiency.

In the feature extraction stage, DilatedLoc is divided into two parts. The first part, Coarse-Grained Sampling, consists of $n=1$ DCD, and the second part, Multi-Scale Context Aggregation, is composed of $n=3$ DCDs.

\begin{figure}[htbp]
\centering
\includegraphics[width=0.75\linewidth]{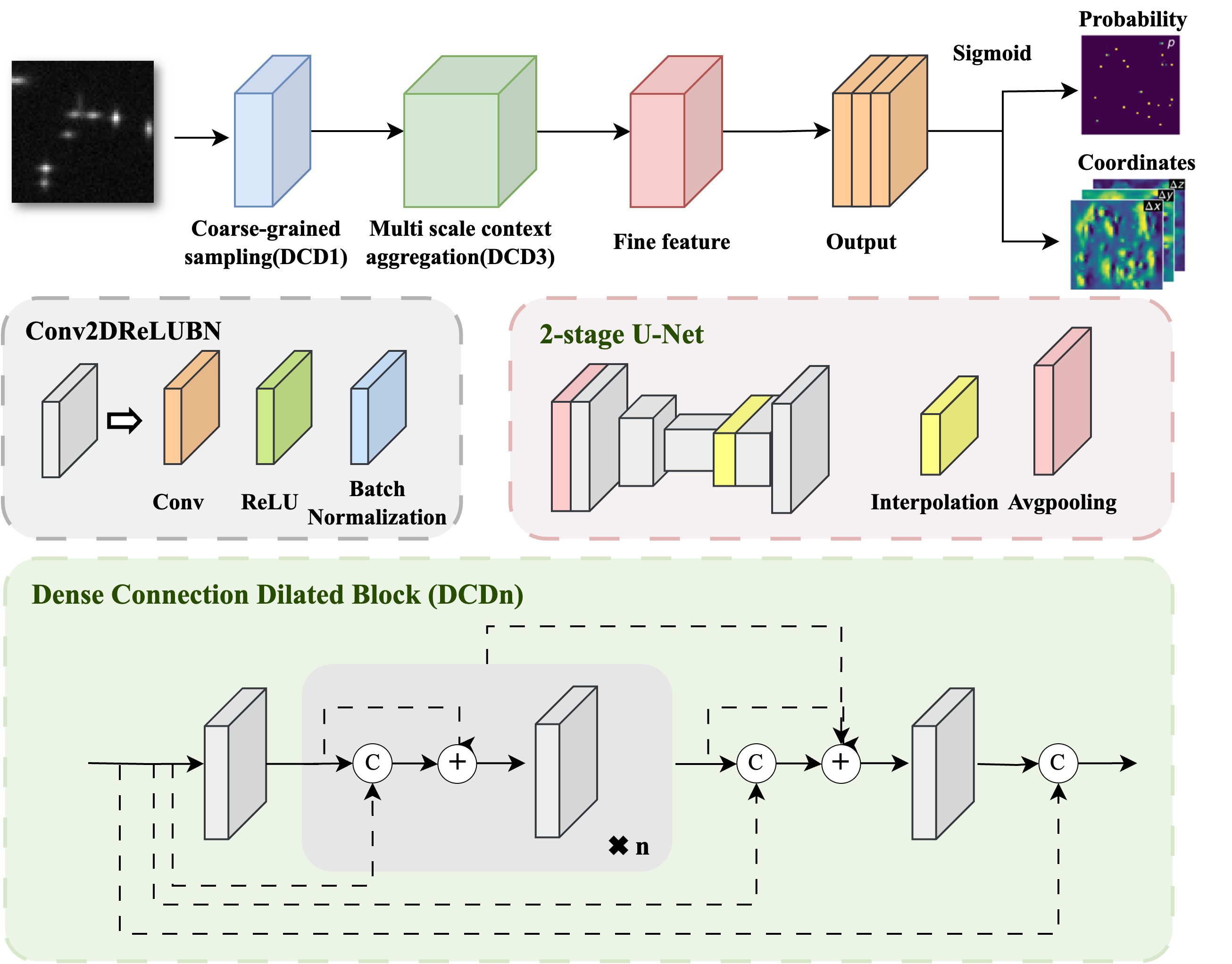}
\caption{The structure of DilatedLoc }
\label{structure}
\end{figure}

After feature extraction, we perform fine-grained processing using a two-stage U-Net \cite{ronneberger2015u}, which maximizes feature extraction while keeping parameters compact. For the network output, we draw on the design from the DECODE network \cite{speiser2021deep}, where each pixel outputs the probability of a fluorescent molecule, along with the offset and uncertainty of the fluorescent molecule’s position relative to that pixel. This design increases both the interpretability and resolution of the network. Detailed network structure information is shown in Figure \ref{structure}.

By combining Dense Connection and dilated convolution, DilatedLoc improves the accuracy of SMLM, especially in high-density conditions.

\subsubsection{Dense Connection Dilated Block}
In high-density single-molecule localization, one of the primary challenges is the overlap of single molecules and the impact of background noise. To address this, our approach uses the structure of DenseNet\cite{jegou2017one} as the core of the network. Dense connections allow each layer to utilize outputs from all preceding layers for feature extraction and classification, significantly enhancing parameter-sharing efficiency.

In constructing the DilatedLoc network, I incorporate dilated convolution as the core element. The key feature of dilated convolution is its ability to control the dilation rate, thereby adjusting the size of the receptive field. By manipulating the dilation rate, the network can effectively combine global and local information, improving the completeness and accuracy of information reconstruction.

One of the most significant advantages of dilated convolution is its impact on z-axis localization accuracy. Standard PSFs often struggle to pinpoint specific locations along the z-axis due to their inherent limitations, which is one of the main reasons for the development of PSF engineering. For instance, astigmatic PSFs artificially introduce aberrations to capture changes near the focal point, encoding z-axis information into their shape. This helps improve z-axis resolution and distinguish positional information along the z-axis. PSFs with a larger axial range tend to have larger shapes \cite{shechtman2015precise} and substantial aberration variations, which traditional convolution methods struggle to capture, making training more difficult.

Dilated convolution is particularly well-suited to solving this problem. By flexibly adjusting the receptive field, it captures the distinctive shapes of PSFs at different positions, enabling more precise localization.

By combining dense connections with dilated convolutions, we create the Dense Connection Dilated Block within the network architecture, effectively enhancing feature collection for both local and global contexts. This design improves the network’s ability to accurately localize single molecules, even in complex, high-density scenarios.

\subsection{Loss Function}

Building on DECODE \cite{speiser2021deep}, we added a cross-entropy function, resulting in a new loss function designed for counting, detecting, and localizing discrete sets of point-like objects.

Cross-entropy is used to measure the difference between two probability distributions and is commonly applied in classification tasks. The value of cross-entropy is always greater than or equal to zero, and it reaches zero only when the predicted values perfectly match the true labels. This loss function provides an intuitive measure of how well the model’s predictions align with true labels, making it an effective guide for the training process.

In multi-class classification problems, where there are $n$ classes, the cross-entropy formula is as follows:$CrossEntropy(y,\hat{y}) =-\sum_{i=1}^n y_i \cdot \log (\hat{y_i}) $where $y_i$ is the probability (0 or 1) of the $i$th class in the true label, $\hat{y_i}$is the probability of the $i$th class predicted by the model. Since predicting the location of fluorescent molecules is actually a binary classification problem, the formula of cross-entropy is as follows: if the true label is y and the predicted label is $\hat{y}$, the final cross-entropy loss function is $loss_{CE} =-(y\log (\hat{y}) +  (1-y)\log(1-\hat{y})$.

In this paper, the cross-entropy function is used to measure the probability of the existence of fluorescent molecules at different positions. When the value of the position is close to 1, it is considered that there is a maximum probability of fluorescent molecules, and when it is close to 0, it is considered that there is no fluorescent molecules at the position. The cross-entropy loss function and the counting loss function together constrain the whole distribution to match the position of the fluorescent molecules, which can effectively guide the training process of the simulation.

\subsection{Ablation Experiments}
To test the impact of dilated convolutions and cross-entropy on the DilatedLoc network, we performed ablation experiments under conditions of high signal-to-noise ratio and low density. We selected 3 $\mu$m Tetrapod PSF images for simulation and conducted a series of tests, outlined in Table \ref{abtext}.

In the first set of experiments, we used the baseline network structure with regular convolutions and the original loss function. Next, we replaced the regular convolutions with dilated convolutions (Baseline+), and finally, we added the cross-entropy function to create a new loss function (Baseline++).

\begin{table}[h!]
\begin{center}
  \caption{Ablation experiment on different condition}
\label{abtext}
\begin{tabular}{cccc}
   \toprule
   Model & Jaccard & RMSE lateral &RMSE axial \\
   \midrule
Baseline &	0.621 &	61.053 &	133.137 \\
Baseline+ &	0.918 &	\textbf{13.092} &	 28.977 	\\
Baseline++ &	\textbf{0.934} &	14.352	&\textbf{ 26.977 }	 \\
   \bottomrule

\end{tabular}
\end{center}
\end{table}

The results demonstrated that dilated convolutions allow for the simultaneous capture of both local and global features, improving information reconstruction and enhancing the network's localization accuracy. When the cross-entropy function was added, the Jaccard Index increased significantly, and axial resolution showed marked improvement, confirming the effectiveness of our approach.

\section{Experiments}
\subsection{Performance of DilatedLoc}
\subsubsection{Cramer-Rao Lower Bound}

\begin{figure}[htbp]
\centering
\includegraphics[width=0.75\linewidth]{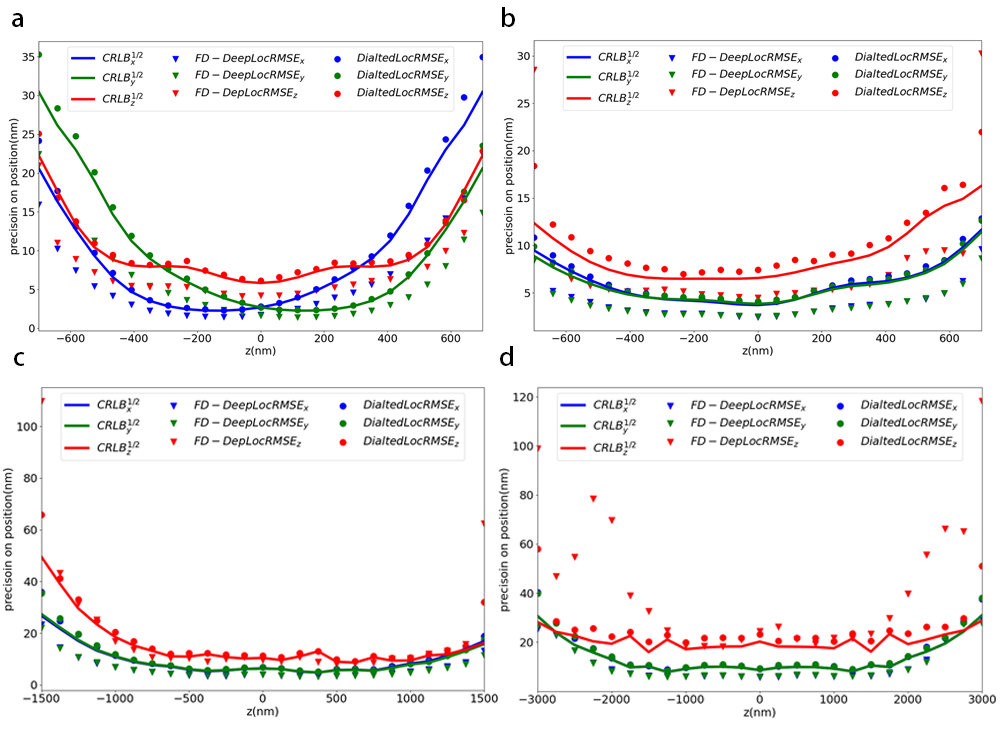}
\caption{DilatedLoc, Fd\_DeepLoc compare the positioning accuracy of four point spread functions with CRLB. a. Astigmatism point spread function; b. Saddle point spread function; c.3$\mu$m tetrapod point spread function; d.6$\mu$m tetrapod point spread functions}
\label{crlbf}
\end{figure}

The Cramer-Rao Lower Bound (CRLB) is a theoretical lower bound that represents the minimum variance of any unbiased estimator. In the context of point spread function (PSF) engineering, CRLB is used to evaluate the quality of PSFs. In this paper, we conducted experiments under conditions of constant brightness and background noise, where we uniformly selected 25 positions along the z-axis and sampled 3,000 points at each position to create a testing dataset. Using this dataset, we compared the localization performance of the DilatedLoc and Fd\_DeepLoc networks, benchmarking their results against the CRLB to evaluate the localization accuracy of both models (see Figure \ref{crlbf}).

We assessed the single-point localization accuracy of four different engineered PSFs with varying axial sizes. As shown in Figure \ref{crlbf}, both Fd\_DeepLoc and DilatedLoc performed relatively close to the CRLB for all four PSFs. For the astigmatic PSF and Saddle Point PSF, Fd\_DeepLoc achieved superior localization accuracy, closely approaching the CRLB. However, as the axial range increased, the localization accuracy of Fd\_DeepLoc began to decline. In contrast, DilatedLoc was closer to the CRLB on the 6$\mu$m Tetrapod PSF, demonstrating its effectiveness in larger axial ranges. The comparison of DilatedLoc’s single-point localization accuracy across different axial ranges shows that this network is capable of performing well with various PSFs.

\subsubsection{SMLM chanage}
By comparing these results with the CRLB, it is evident that DilatedLoc has strong localization performance, comparable to other high-accuracy networks. To further validate the localization capability of DilatedLoc, we conducted a comparison with Fd\_DeepLoc and DECODE using the public dataset from the 2016 SMLM Challenge.

The 2016 SMLM Challenge is a second-generation, continuously updated comprehensive benchmark designed to objectively and quantitatively evaluate the wide array of available localization algorithms. It provides synthetic datasets for training, simulating various experimental conditions. To prevent overfitting, the evaluation is performed on data that is not shared with the participants. The challenge computes various quality metrics, including root mean square error (r.m.s.e) for 2D and 3D data, JI for detection accuracy, and a single "efficiency" score that combines r.m.s.e and JI.

Based on the fluorescence blinking positions and intensities provided by the challenge, we selected different PSF models for data simulation, resulting in four sets of data. By comparing the performance metrics of these three networks on the public dataset (see Table \ref{smlmaa}), we observed that the DilatedLoc model exhibited slightly lower localization accuracy than Fd\_DeepLoc for the astigmatic PSF model. However, as the PSF size increased, DilatedLoc's performance gradually surpassed that of Fd\_DeepLoc, with particularly outstanding results for the Tetrapod PSF.

The results show that while DilatedLoc slightly underperforms Fd\_DeepLoc on smaller PSFs, such as the astigmatic PSF and Saddle Point PSF, it demonstrates significantly better localization performance on the 3$\mu$m and 6$\mu$m Tetrapod PSFs, particularly in terms of root mean square error (RMSE). Dilated convolution enhances both local and global feature extraction by adjusting the receptive field, an advantage that becomes more evident with larger PSFs. This trend is consistent with the single-point localization results.

\begin{table}[htbp]
\caption{Performance of DilatedLoc and Fd\_DeepLoc in the challenge with high SNR and low density}
\label{smlmaa}
\centering
 \begin{tabular}{ccccc}
\toprule
PSF &Model & Jaccard& RMSE Lateral & RMSE Axial \\
 \midrule
Astigmatism& FD-DeepLoc & \textbf{0.980} &18.621& \textbf{20.743}\\
&DilatedLoc & 0.975 &\textbf{14.614}& 21.33\\
 \midrule
DMO-SaddlePoint& FD-DeepLoc & 0.966 &17.577& \textbf{23.234}\\
&DilatedLoc & \textbf{0.978} &\textbf{11.884}& 23.559\\
 \midrule
Tetrapod( 3$\mu$m)& FD-DeepLoc & \textbf{0.967 }&21.218& 36.929\\
&DilatedLoc & 0.965 &\textbf{12.858}& \textbf{29.068}\\
 \midrule
Tetrapod( 6$\mu$m)& FD-DeepLoc & 0.910 &26.019& \textbf{46.948}\\
&DilatedLoc & \textbf{0.916} &\textbf{15.153}& 34.022\\
   \bottomrule

\end{tabular}
\end{table}

Through ablation experiments and cross-network comparisons, we conclude that dilated convolution provides better feature capture for larger axial PSFs. By adjusting the dilation rates for PSFs of different sizes, the network achieves more precise localization. The current dilation parameters used in this paper are (2, 4, 8, 16), and tests show that this parameter set delivers good performance across all four PSF types, with particularly excellent results for large axial PSFs, achieving theoretical localization accuracy.

\subsubsection{High resolution on Large Axial PSFs}

 In experimental conditions, the overlap of single-molecule points is inevitable, significantly affecting the performance of localization algorithms. In almost all applications, there is a need to simultaneously localize nearby single molecules. For instance, in super-resolution SMLM experiments, the number of fluorescent molecules localized per frame determines the time resolution. In tracking applications, overlapping PSFs from multiple fluorescent molecules can hinder localization and reduce accuracy, leading to large localization errors in densely packed regions.

This issue is particularly pronounced in three-dimensional (3D) localization. Specifically, in large axial ranges (greater than 3 $\mu$m), encoding the axial position of a single molecule requires using laterally larger PSFs, such as the Tetrapod PSF, which increases the likelihood of overlap. Current algorithms that perform well in 3D high-density localization, including DeepSTORM3D, DECODE, and Fd\_DeepLoc, face challenges in these dense environments.

\begin{figure}[h!]
\centering
\includegraphics[width=0.75\linewidth]{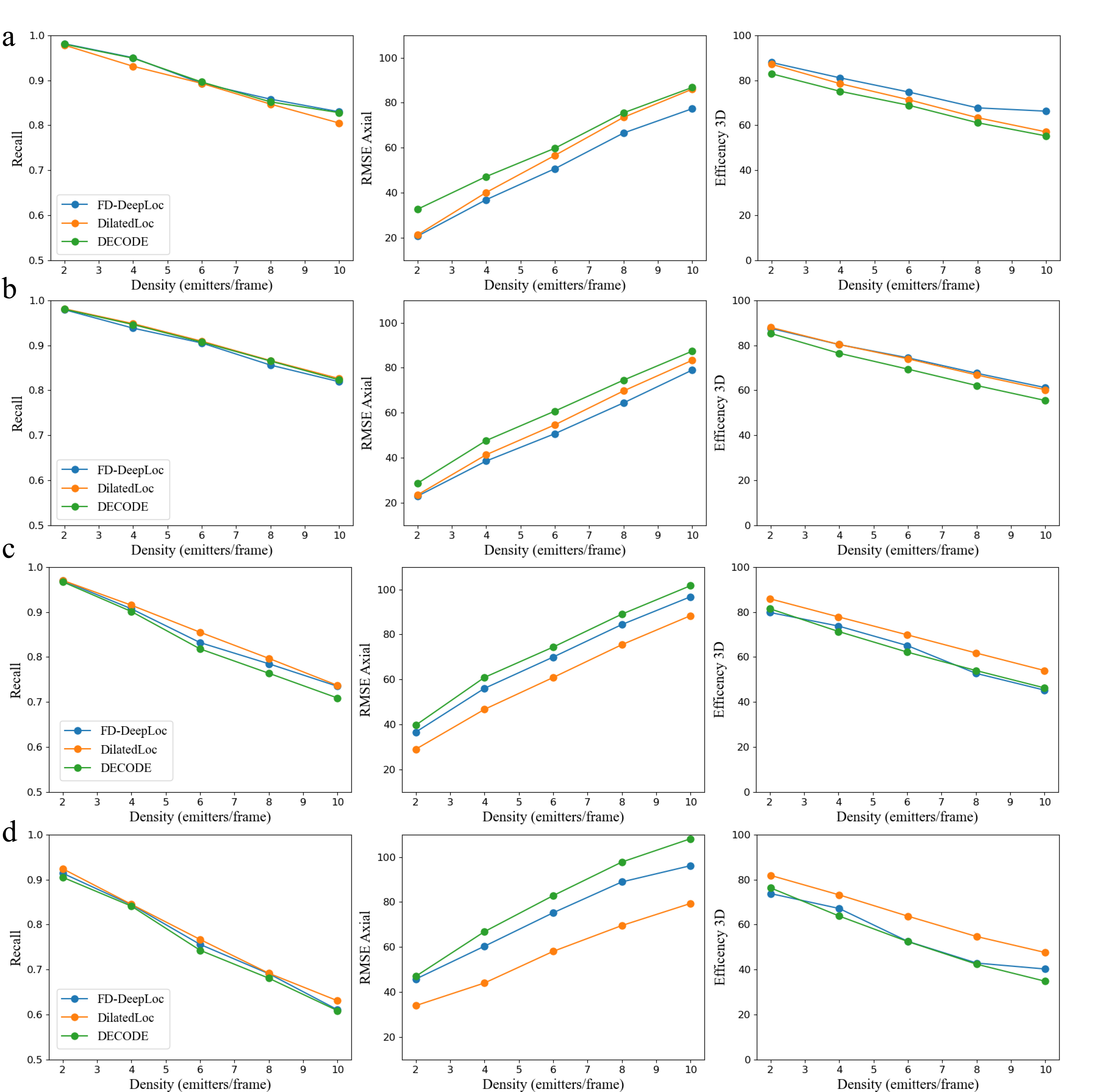}
\caption{DilatedLoc,Fd\_DeepLoc, DECODE evaluation results on images generated by four point spread function simulations. The figure shows three metrics, namely recall, root mean square error, 3D efficiency value. a. astigmatism point spread function; b. saddle point spread function; c.3$\mu$m quadruped point spread function; d.6$\mu$m quadruped point spread function}
% \caption{performance}
\label{self_dense}
\end{figure}
However, the DilatedLoc network consistently shows robust performance across different densities (Figure \ref{self_dense}). As the density increases, the likelihood of PSF overlap grows, making localization increasingly difficult. We found that higher densities lead to varying degrees of performance degradation. The metric most affected by density is the recall rate, particularly for larger PSFs, where performance declines more significantly (Figure \ref{self_dense}d).

The cross-entropy loss function used in the DilatedLoc network improves precision (P), ensuring higher accuracy in prediction results. The increase in density also affects root mean square error (RMSE), as demonstrated in the figures. As the density increases and the probability of overlap rises, all three models exhibit some degree of RMSE degradation. For the astigmatic PSF (Figure \ref{self_dense}a), the performance drop in DilatedLoc (orange line) with increasing density is more pronounced compared to Fd\_DeepLoc (blue line). However, as the axial range increases, DilatedLoc’s accuracy becomes more stable, showing slower degradation than Fd\_DeepLoc as density increases. Therefore, for large axial PSFs, DilatedLoc maintains better stability and higher localization accuracy under increased density. The final column in Figure \ref{self_dense} represents the Efficiency Volume (Equation \ref{effic} \cite{sage2019super}), a single metric to evaluate localization performance.

\begin{equation}
\begin{aligned}
\operatorname{Efficiency_{volume}} &= 0.5 \times (\operatorname{Efficiency_{lateral}} + \operatorname{Efficiency_{axial}})\\
& = 0.5 \times \Bigg(1 - \sqrt{(1-\operatorname{Jaccard\ Index})^{2} +  r.m.s.e._{lateral} ^{2}} \\ & + 1 -\sqrt{(1-\operatorname{Jaccard\ Index})^{2} + 0.25\times r.m.s.e._{axial} ^{2}} \Bigg)\\
\end{aligned}
\label{effic}
\end{equation}

Figure \ref{self_dense}b shows the performance of the Saddle Point PSF across the three networks as density increases. In terms of recall rate, all three networks perform comparably. In axial resolution, Fd\_DeepLoc delivers the highest localization accuracy, followed by DilatedLoc and DECODE. The Efficiency3D metric follows a similar trend. In Figure \ref{self_dense}c, which evaluates the 3$\mu$m Tetrapod PSF, recall rates drop across all networks as density increases, with DECODE being most affected, while DilatedLoc is least impacted. In terms of axial resolution, DilatedLoc shows a clear advantage, maintaining an RMSE below 80. 

In Figure \ref{self_dense}d, the results for the 6$\mu$m Tetrapod PSF mirror those of the 3$\mu$m Tetrapod PSF. DilatedLoc consistently demonstrates superior localization performance, and its advantage over the 3$\mu$m Tetrapod PSF is even more pronounced.

The results clearly show that DilatedLoc performs best for the 6$\mu$m Tetrapod PSF, followed by the 3$\mu$m Tetrapod PSF, with the Saddle Point and astigmatic PSFs ranking third and fourth, respectively. This trend holds across varying densities. Thus, we conclude that the DilatedLoc network is well-suited for large axial PSFs, offering high localization accuracy regardless of density conditions.

\subsection{Fast inference Imaging}
\subsubsection{Contrast Experiment}
For testing, we selected 20,000 image frames (see Table \ref{timet}) and performed the computations on an RTX 3080 GPU with a batch size of 10. FLOPs (floating point operations per second) refers to the number of floating point calculations and is commonly used to measure the complexity of an algorithm or model.
\begin{table}[ht]
\begin{center}
  \caption{Inference time comparison on DilatedLoc and FD\_Deeploc}
\label{timet}
\begin{tabular}{cccc}
   \toprule
    & DilatedLoc & FD\_DeepLoc & Ratio\\
   \midrule
   Parameter Number& \textbf{453835} & 2522746  & 5.55 \\
   FLOPs(128 $\times$ 128)& 2.024E+10 & \textbf{1.622E+10} & 0.8\\
   time cost(64$\times$64)ms/per frame& \textbf{1.14} & 1.58 & 1.38 \\
   time cost(128$\times$128)ms/per frame&\textbf{ 3.205} & 5.948 & 1.855\\
   \bottomrule
\end{tabular}
\end{center}
\end{table}

From the results in Table \ref{timet}, it is evident that inference speed correlates with image size—larger images result in increased inference time. When the image size was 64x64, FD\_DeepLoc required 1.38 times the inference time of DilatedLoc. For an image size of 128x128, the inference time for FD\_DeepLoc increased to 1.8 times that of DilatedLoc.

Although FD\_DeepLoc has fewer FLOPs than DilatedLoc, its overall inference time per frame and parameter count are significantly higher. There are three primary reasons for this:
1. FD\_DeepLoc is built on a U-Net architecture that downsamples input at different stages, reducing the width and height, thus reducing computational requirements. However, this makes FD\_DeepLoc more sensitive to image size, increasing inference time for larger images.
2. DilatedLoc, while having fewer layers, uses two upsampling stages to increase input size during fine-grained processing, which increases computational complexity despite having fewer parameters than FD\_DeepLoc.
3. The longer inference time for FD\_DeepLoc is also partly due to its post-processing requirements. Since the model outputs offsets for the \(x, y, z\) coordinates, extra time is needed to convert these offsets into absolute coordinates. In contrast, DilatedLoc has been optimized to reduce post-processing time, despite using a similar output format.

\subsubsection{Inference frame performance experiment}

\begin{figure}[h!]
\centering

\includegraphics[width=0.5\linewidth]{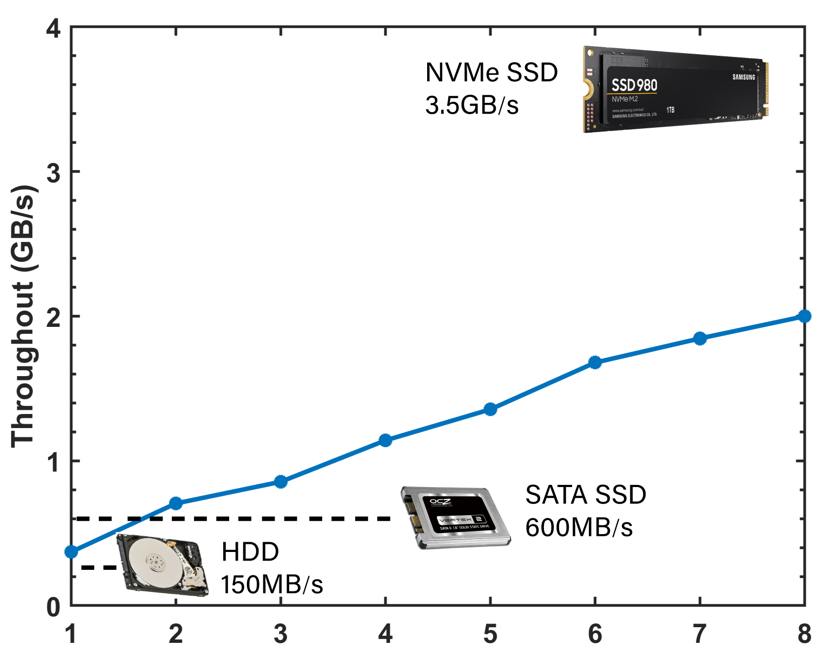}
\caption{Performance of throughput with the increase of the number of GPU devices }
\label{speed}
\end{figure}

Our prior tests and the results presented in the paper show that single-molecule image density does not significantly impact the inference time for individual frames. To assess the multi-GPU performance, we randomly simulated 200,000 frames of 128x128 float32 single-molecule random dot images as the test dataset. Figure \ref{speed} shows the throughput results obtained using varying numbers of GPUs. The tests were conducted on an NVIDIA GeForce RTX 4090 with NVME SSD storage, a batch size of 128, and GPU utilization consistently exceeding 95\% throughout inference.

From the figure, it is clear that the throughput with a single GPU exceeds the read speed of traditional mechanical hard drives. With two GPUs, the throughput surpasses that of SATA SSDs, reaching approximately 600 MB/s. When using 8 GPUs in parallel, throughput approaches 2 GB/s.

\section{Discussion}

The proposed framework achieves high throughput, ensuring optimal GPU performance even when processing large images at high inference speeds. This results in significant improvements in overall inference speed. Additionally, DilatedLoc, designed using dilated convolution and dense connections, is well-suited for high-density localization tasks over large axial ranges. Despite having only one-third of the parameters of FD\_DeepLoc, DilatedLoc still delivers high localization accuracy. Due to the Dense Connection Dilated Block (DCD), the network excels in localizing large axial PSFs with high precision. DCD efficiently combines local and global features, improving localization performance, particularly for larger axial PSFs.

However, for smaller PSFs, such as the astigmatic PSF, the advantages of DilatedLoc are less pronounced. While it maintains decent localization accuracy, its performance under high-density conditions does not surpass that of FD\_DeepLoc.

By integrating DilatedLoc with our proposed inference framework, we further leverage its fast imaging capabilities. Rapid imaging is crucial in live-cell SMLM, where the dynamic nature of biological systems demands high time resolution. Typically, fast imaging requires high-power lasers, which can reduce resolution and introduce severe phototoxicity. DilatedLoc’s ability to precisely localize in high-density conditions allows for higher activation of fluorescent molecules, meaning faster imaging can be achieved with lower light doses. This makes DilatedLoc a valuable tool for high-throughput imaging in live-cell environments, offering critical algorithmic support for such applications.

\bibliographystyle{IEEEtran}
% \bibliography{references}

\end{document}